
\magnification=1200
\baselineskip20pt
\def\DM{\displaystyle}

\hglue4in
UM-TH-93-20
\smallskip
\hglue4in
IP-ASTP-29-93
\smallskip
\hglue4in
August 25, 1993
\bigskip
\hglue1.4in
{\bf Exact $\alpha _s$ Calculation of
$b\rightarrow s + \gamma $}
\smallskip
\hglue2.2in
{\bf $b\rightarrow s + g$}
\bigskip
\hglue1.5in
{\bf K. Adel$^a$  \ and \  York-Peng Yao$^{a,b}$ }

$$^a \ Department \ of \ Physics, \ Randall \ Laboratory, \ University \ of \
Michigan,$$
$$Ann \ Arbor, \ Michigan \ 48109-1120, \ U.S.A.$$
$$^b \ Institute \ of \ Physics, \ Academia \ Sinica, \ Taipei, $$
$$Taiwan \ 11529, \ Republic \ of  \ China.$$
\bigskip
\noindent
{\it Abstract:}
\medskip
We present an exact $\alpha_s$ calculation of the Wilson coefficients
associated with the dipole moment operators. We also give an estimate
of the branching ratio for $b\rightarrow s \gamma$. We find that higher
dimensional effects are under control within $9\%$ for
$BR(b\rightarrow s \gamma)=(4.3\pm 0.37  )\times 10^{-4}$.

\vglue1in
\noindent
PAC: 11.10Jj, 12.38.Bx, 13.40.Hq

\vfill \eject

In an article$^1$ by us recently, we presented a leading logarithmic
analysis of the
heavy particle effects on the process $b \rightarrow s \gamma $, which
incorporates a complete operator mixing.
Among the results, we find that our mixing matrix differs in some
elements from the work by Misiak$^2$.  Had we subscribed to the prescription
of further mixing with evanescent operators$^3$, we would have obtained extra
contributions (in unit of $g_s^2/8\pi ^4$)
$$\Delta \gamma _{O_{67} \ O_{51}}=1/3, \ \ \Delta \gamma _{O_{67} \ O_{52}}=
-8/3,$$
$$\Delta \gamma _{O_{68} \ O_{51}}=-1/2, \ \ \Delta \gamma _{O_{68} \ O_{52}}=
-4.\eqno (1)$$
These would have made the mixing by Ciuchini et. al.$^4$ coincide with ours.
(We thank Misiak for correspondence on this.)  For earlier work, see$^{5,6}$.
\bigskip
More importantly, by choosing two
different limits for extrapolation: $m_t = m_w$ and $m_t \gg m_w$,
we estimated the effects due to higher orders in $m_w^2/m_t^2$ to be
about $20\%.$  In view of recent interests in the experimental branching
ratio$^7$ for $B_s \rightarrow K^\star +\gamma $ and an impending value for the
inclusive rate $B_s \rightarrow X_s + \gamma $, together with the
sensitivity of these processes as a short distance probe, an uncertainty
in short distance analysis of $20\% $ is hardly satisfactory or even
acceptable.  The purpose of this publication is to provide some
remedy.
\bigskip
One can trace the uncertainty to the fact that $m_w^2/m_t^2 \approx 25 \%$ for
$m_t \approx 150$ Gev. It strongly suggests that one should calculate the order
$\alpha_s$
diagrams exactly, whereever there are internal top and/or W-boson propagators.
This is what we have done.
\bigskip
In all two-loop diagrams (and their attendant counter terms)
which contribute to this calculation, we keep all orders in
$x \equiv m_t^2/m_w^2$ but
discard terms proportional to $O(m_b^2/m_{t,w}^2)$ or $O(m_s^2/m_{t,w}^2)$
(we have factored out the Fermi weak coupling constant).
The amount of algebra is highly non-trivial and is aided by Schoonschip.$^8$
\bigskip
Given a diagram in which there are  top and/or  $W-\phi $ internal lines,
there are various sequences of operations one can follow to isolate its
dependence on the heavy masses$^9$.  In any case,
the underlying method is based on partitioning of the diagrams into heavy parts
and operator inserted matrix elements.
For the present
situation, we treat the top and the $W-\phi $ as being correspondingly
heavy, relative to other masses and external momenta.  In the order we
are working with, a heavy part
always contains both the top and the $W-\phi $ internal lines.
Vertices made
of light particles and momenta acting on them and Wilson coefficients
which contain all dependence of heavy masses are organized within this
formalism. In this way, we obtain the effective Lagrangian
$$L_{eff}=\sum_iC_iO_i, \eqno (2)$$
where $C_i$ are the Wilson coefficients and $O_i$ are a set of local
operators, made of light fields.
\bigskip
Of particular interest in the present order $\alpha _w \alpha _s$ calculation
are the coefficients $C_{O_{51}}$ and $C_{O_{52}}$, with the accompanying
operators
$$O_{51}=ig_s\bar sG_{\mu \nu }\sigma _{\mu \nu }(m_sP_L+m_bP_R)b/2,\eqno (3)$$
and $O_{52}=O_{51}(g_s G_{\mu \nu}\rightarrow (-1/3)eF_{\mu \nu}).$
\bigskip
Before we go on further, let us define
$$\bar C_{O_{5i}}\equiv 16\pi ^2C_{O{5i}} /G_t \ (i=1,2), \ \ {\rm and} \ \
G_t=2\sqrt 2G_FV^\star_{ts}V_{tb},\eqno (4)$$
where $G_F$ is the Fermi weak coupling constant and $V$'s are the
Cabbibo-Kobayashi-Maskawa matrix elements.  Then the exact $\alpha _s$ results
are (in the $R^*$ scheme to be defined)
$$\eqalign{ &{\pi \over 4\alpha _s} \bar C^{(1)}_{O_{51}} =
  - {4\over (x-1)^4}
  - {5293\over 576}\,{1\over (x-1)^3}
  - {21989\over 3456}\,{1\over (x-1)^2}
  - {1817\over 1728}\,{1\over (x-1)}
  + {247\over 10368}
\cr
& + \log\left({m_w^2\over \mu^2}\right) \, \left[\
   - {7\over 16}\,{1\over (x-1)^3}
   - {21\over 32}\,{1\over (x-1)^2}
   - {7\over 48}\,{1\over (x-1)}
   - {35\over 288}
\ \right]  \cr
& + \log\left({m_w^2\over \mu^2}\right)\,\log x  \, \left[\
    {7\over 16}\,{1\over (x-1)^4}
   + {7\over 8}\,{1\over (x-1)^3} +{7\over 16}\, {1\over (x-1)^2} \ \right]
\cr
& + \log x    \left[
	{4\over (x-1)^5}
	+ {5509\over 576} {1\over (x-1)^4}
	+ {2893\over 432}  {1\over (x-1)^3}
	+ {163\over 192} {1\over (x-1)^2}
        - {11\over 48} {1\over (x-1)}   - {35\over 288}
\right] \cr
& + \log^2 x  \, \left[\
	+ {7\over 16}\,{1\over (x-1)^4}
	+ {7\over 8}\,{1\over (x-1)^3}
	+ {7\over 16}\, {1\over (x-1)^2}  \ \right] \cr
& + {\rm Sp}\left(1-{1\over x}\right) \, \left[\
	 {13\over 8}\,{1\over (x-1)^4}
	+ {187\over 48}\,{1\over (x-1)^3}
	+ {137\over 48}\,{1\over (x-1)^2}
	+ {1\over 2}\,{1\over (x-1)} - {1\over 12}  \ \right] \cr
& +\left( {91\over 2592} + {1\over 54} \log \left( {m_w^2\over \mu^2} \right)
\right) ,   \cr        } $$

$$\eqalign{ &{\pi\over 4\alpha _s} \bar C^{(1)}_{O_{52}} =
	- {4\over (x-1)^4}
	- {335\over 18}\,{1\over (x-1)^3}
	- {11959\over 432}\,{1\over (x-1)^2}
	- {6347\over 432}\,{1\over (x-1)} - {47\over 648} \cr
& + \log\left({m_w^2\over \mu^2}\right)  \, \left[\
	- {1\over (x-1)^3} -{3\over (x-1)^2}
	- {31\over 12}\,{1\over (x-1)} + {17\over 36} \ \right] \cr
& + \log\left({m_w^2\over \mu^2}\right)\,\log x  \, \left[\
	 {1\over (x-1)^4}
	+ {7\over 2}\,{1\over (x-1)^3} +{4\over (x-1)^2}
        + {3\over 2}\,{1\over (x-1)}  \ \right] \cr
& + \log x    \left[\
	{4\over (x-1)^5}
	+ {643\over 36} {1\over (x-1)^4}
	+ {1337\over 54} {1\over (x-1)^3}
	+ {257\over 24} {1\over (x-1)^2}
	- {19\over 24} {1\over (x-1)} + {17\over 36}  \ \right] \cr
& + \log^2 x \, \left[\
	 {1\over (x-1)^4}
	+ {7\over 2}\,{1\over (x-1)^3} +{4\over (x-1)^2}
	+ {3\over 2}\,{1\over (x-1)}  \ \right] \cr
& +{\rm Sp}\left(1-{1\over x}\right)  \, \left[\
	{11\over 4}\,{1\over (x-1)^4}
	+ {139\over 12}\,{1\over (x-1)^3} + {191\over 12}\,{1\over (x-1)^2}
	+ {31\over 4}\,{1\over (x-1)} + {2\over 3} \ \right]   \cr
& +\left( -{713\over 648}+{1\over 54}\log\left({m_w^2\over \mu^2}\right)
\right) , \cr }\eqno (5) $$
where the last terms in big parenthesis in each of the two equations above are
the contributions due to $u$ and $c$ quarks. Sp stands for Spence function.
Eq.(5) are to be contrasted with the asymptotic results ($m_t\gg m_w$)
$$\bar C^{(1)}_{O_{51}}(asym)\pi /4\alpha =
 {611\over 10368}-{\pi^2\over 72 }
-{35\over 288}\log\left({m_t^2\over \mu^2}\right)
+{1\over 54}  \log\left({m_w^2\over \mu^2}\right),$$
and
$$\bar C^{(1)}_{O_{52}}(asym)\pi /4\alpha =
-{95\over 81}+{\pi ^2\over 9}
+{17\over 36} \log\left({m_t^2\over \mu^2}\right)
+{1\over 54}  \log\left({m_w^2\over \mu^2}\right),
\eqno (6)$$
obtained by discarding all $O(m_w^2/m_t^2).$  One can use Eq.(5) to check
some of the mixing \hbox{matrix} elements of the renormalization
group equations
(RGE).  Unfortunately, $\gamma _{O_{67} \ O_{51}}$,
$\gamma _{O_{67} \ O_{52}}$, $\gamma _{O_{68} \ O_{51}},$ and
$ \gamma _{O_{68} \ O_{52}} $ do not
enter to this order.  Their determination has to come from three
loop diagrams. In our opinion, direct
Feynman diagram computation of Green's functions for processes to extract
mixing matrix elements should be the definitive procedure.
\bigskip
We plot the exact $\alpha _s$ result Eq.(5) in Figure 1, together with the
asymptotic result Eq.(6) for $\bar C^{(1)}_{O_{52}}\pi /4 \alpha _s$. For
x=4, the discrepancy is in fact about $50\% $.  It is interesting to note that
the exact $C^{(1)}_{O_{52}}$ is quite flat between x=1 to x=10.
The dotted line is the $\alpha _s^0$ exact result,
also known as the Inami-Lim functions$^{10}$
$$\bar C^{(0)}_{O_{51}}={-2x-3x^2+6x^3-x^4-6x^2\log x \over 4(1-x)^4},$$
$$\bar C^{(0)}_{O_{52}}={7x-12x^2-3x^3+8x^4+(12x^2-18x^3)\log x\over
4(1-x)^4}.\eqno (7)$$
We see that below x=6.9, second order QCD correction is bigger
than the lowest order result.  This has been known for the approximate
results for some time and in fact is an impetus for looking into rare
decays of this genre.
\bigskip
Please note that if one uses Eq.(7) as (a part of) the boundary conditions,
then Eq.(5) are the $\alpha _s $ solution of RGE for all values of $x$, in so
far as the overall coefficients to $\log \mu ^2$ are concerned. We would
like to stress that we are using the $R^*$ scheme.  In
the $\overline{MS}$ scheme, the exact result to order $\alpha_s$ is
obtained if one makes the replacement $m_t \rightarrow
\DM m_t \left( 1 - \alpha_s/\pi\;\log(m_t^2/\mu^2)\right)$ in Eq. (7);
this replacement is just a finite renormalization to go from the $R^*$ scheme
to the $\overline{MS}$ scheme.
\bigskip
We shall make the assumption that after $\alpha_s $ corrections, it is
safe to add the leading logarithmic terms to complete the leading QCD sum.
In other words, we assume that the higher order QCD corrections can
be obtained either in the limit $m_w^2/m_t^2 \ll 1$ or in the limit
$m_w^2/m_t^2 = 1$.
This assumption can be tested as in our previous publication.
For $\bar C_{O_{52}}$, there entail two different extrapolations
$$\bar C^{(0)+(1)}_{O_{52}}(exact)+\bar C_{O_{52}}^{higher \ order}(m_t=m_w)
\eqno (8.a),$$
and
$$\bar C^{(0)+(1)}_{O_{52}}(exact)+\bar C_{O_{52}}^{higher \ order}(m_t\gg m_w)
,\eqno (8.b)$$
where $\bar C_{O_{52}}^{higher \ order}(m_t=m_w)$ and
$\bar C_{O_{52}}^{higher \ order}(m_t\gg m_w)$ are the remaining leading
logarithmic
sums with the boundary conditions set at $m_t=m_w$ and $m_t\gg m_w$,
respectively.$^1$
\bigskip
For the physical process $b\rightarrow s +\gamma$, some four quark operators
also contribute, resulting in an effective coupling$^2$
$$C^{eff}_{O_{52}}=C_{O_{52}}+{1\over 8\pi ^2}C_{O_{67}}+{3\over 8\pi ^2}
C_{O_{68}}.\eqno (9)$$
Also, to remove the dependence on $|G_t|^2$, which is not accurately known
expermientally, we normalize the $b \rightarrow s \gamma $ partial width to
the well established semileptonic $b \rightarrow ce\bar \nu$ partial width, and
use the following relation$^{11}$ $|V^\star_{ts}V_{tb}|\simeq |Vcb|$.

This ratio is given as:
$$ {\Gamma \left(b \rightarrow s \gamma\right)
\over \Gamma \left( b \rightarrow c e \bar{\nu} \right) } \simeq
{ \alpha_{QED} \over 6 \pi \, g\,(m_c/m_b)}\,
\left( 1 - {2\alpha_s(m_b) \over 3 \pi} f(m_c/m_b) \right)^{-1}\,
|\bar {C}_{O_{52}}^{\,eff}(m_b)|^2 ,\eqno (10)$$
where $g(m_c/m_b)\simeq 0.45$ and $f(m_c/m_b)\simeq 2.4$ correspond to
the phase space factor and the one-loop $QCD$ corrections to the
semileptonic decay, respectively.

In Figure 2, we have plotted  $BR \left( b \rightarrow s \gamma\right)$
as a function of $m_t$. The solid and dashed curves are obtained
with the aid of the interpolation equations in Eq.(8),
together with:
$$ BR \left( b \rightarrow s \gamma\right) =
{\Gamma \left(b \rightarrow s \gamma\right)
\over \Gamma \left( b \rightarrow c e \bar{\nu} \right) }
\  BR \left( b \rightarrow c e \bar{\nu} \right), \eqno (11)$$
$$ \  BR \left( b \rightarrow c e \bar{\nu} \right) \simeq 0.108. \eqno (12)$$
\noindent
The vertical dotted line to guide the eyes intersects these curves
at $m_t=140 \ GeV$ and gives respectively:
$$ \eqalign{ & BR \left( b \rightarrow s \gamma\right) =  4.66\;\times 10^{-4},
\cr
             & BR \left( b \rightarrow s \gamma\right) = 3.93\;\times 10^{-4}.
\cr} \eqno (13)$$
\noindent
{}From Eq. $(13)$, we obtain the following mean value (this is our estimate),
$$ BR \left( b \rightarrow s \gamma\right) = (4.3\pm 0.37  )\times 10^{-4}.
\eqno (14)$$
This is to
be compared with an upper limit $5.4\times 10^{-4}$ given recently by the
CLEO Collaboration.$^{(7)}$
\smallskip
The uncertainty due to subleading logarithmic and higher dimensional
effects is about $9\%$, which is a big improvement  and more reliable
over what we gave before, where $m_t^2/m_w^2$ effects
at $\alpha _s ^{(1)}$ were not treated.


\bigskip
\medskip
We now give some technical details. We use the general linear covariant
gauges ${-1\over 2\alpha }(\partial _\mu G_\mu )^2$ for the gluons.  The
complete cancellation of $\alpha $ for $C_{O_{51}} $ and $C_{O_{52}} $ is
a stringent confirmation on the correctness of the algebra.  The gauge fixing
for W-fields is $-C^+C^-$ with $C^+=-\partial _\mu W_\mu^++m_w\phi ^+
+ieA_\mu W_\mu.$
For oversubtractons and renormalization, we use the $R^\star $-scheme.  Thus,
let $\Gamma $ be a one light particle irreducible (1LPI) diagram which
contains the heavy particles, and let $\gamma $ represent a 1LPI
graph or subgraph ($\gamma \subseteq \Gamma$) with external generic momentum p.
We define
$$\tau_\gamma^\epsilon =pole \  part  \ of  \ \gamma , \ \epsilon =n-4,$$
$$\tau_\gamma ^{(m)}=\gamma (p=0)+p{\partial \over \partial p}\gamma (p=0)
+ \cdot \cdot \cdot + {1\over m!}p^m{\partial ^m\over \partial p^m}\gamma
(p=0).\eqno (11)$$
The $R^\star $ renormalization procedure is defined as
$$R^\star (\gamma _{heavy})=(1-\tau_\gamma ^{(m)})\gamma _{heavy},\ \ \
R^\star (\gamma _{light})=(1-\tau_\gamma ^\epsilon)\gamma _{light},\eqno (12)$$
where $m$ is so chosen that the neglected terms are genuinely of
$O(1/m_{heavy}^2)$.
It is important to repeat that the results Eqs.(5,6) are
given in the renormalized top mass under the $R^\star $-scheme,
where $\beta_{m_t}=0.$
\bigskip
Except for trivial factorizable cases, all two loop integrals we need are
related to$^{12}$
$$ \eqalign {
I_{2,1,1} & (m_1^2,m_2^2,m_3^2)  =\int d^nkd^nq
{1\over (k^2+m_1^2)^2((k+q)^2+m_2^2)(q^2+m_3^2)} \cr
& =\pi ^4 \left[ {-2\over (n-4)^2}+{1\over n-4}(1-2\gamma_E-2\log (\pi m_1^2))
-{1\over 2}-{1\over 12}\pi ^2   \right]  \cr
& \ \ +\pi ^4 \left[ \gamma _E -\gamma _E^2+(1-2\gamma _E)\log (\pi m_1^2)
-\log ^2(\pi m_1^2)
+f(a={m_2^2\over m_1^2},b={m_3^2\over m_1^2}) \right] ,
\cr }\eqno (13)$$
where
$$\eqalign {f(a,b)={1\over 2}\log a \log b+{a+b-1\over \sqrt {\ \  }}  &
[{\rm Sp}({-y_2\over x_1})+{\rm Sp}({-x_2\over y_1})+{1\over 4}\log ^2
{x_2\over y_1} \cr
& +{1\over 4}\log ^2 {y_2\over x_1}+{1\over 4}\log ^2{x_1\over y_1} -{1\over 4
}
\log ^2 {x_2\over y_2} + {\pi ^2 \over 6} ] , \cr }\eqno (14)$$
$$x_{1,2}={1\over 2}(1-a+b\pm \sqrt {\ \ }), \ \
y_{1,2}={1\over 2}(1+a-b \pm \sqrt {\ \ }   ),$$
and
$$\sqrt { \ \ }= \sqrt {(1-a+b)^2-4b}.\eqno (15)$$
\bigskip
Assuming $a,b \gg 1$, (i. e. we take
$m_2=m_t, \ m_3=m_w \gg m_1=m_{b,s}$), we expand $f(a,b)$
in series of $m_{b,s}^2$, which must be
retained to proper orders in intermediate steps.
\bigskip
Details of this work are to be published$^1$.
\bigskip
This work has been partially supported by the U. S. Department of Energy.
Y.-P. Y. would like to thank members of the Particle Theory Group at
the Institute of Physics, Academia Sinica, for hospitality.
\vfill
\eject
\noindent
{\bf References:}
\bigskip

\settabs 20 \columns
\+ [1] & K. Adel and York-Peng Yao,  Modern
             Physics Letters A {\bf 8}, 1679 (1993); \cr
\+     & also under preparation.\cr

\medskip

\+ [2] & M. Misiak,  Phys. Lett. {\bf 269 B}, 161 (1991); Nucl.
               Phys. {\bf B393}, 23 (1993). \cr
\medskip

\+ [3] & A. Buras and P. Weisz, Nucl. Phys. {\bf B333}, 66 (1990). \cr

\medskip

\+ [4] & M. Ciuchini, E.Franco, G. Martinelli, L. Reina, L. Silvestrini,
               LPTENS 93/28, \cr

\+     & ROME 93/958, ULB-TH 93/09 (unpublished). \cr

\medskip

\+ [5] & B.~Grinstein, R.~ Springer and M.~B.~Wise,  Phys. Lett.
                 {\bf 202 B}, 138 (1988); \cr
\+     & P.~Cho and B.~Grinstein,  Nucl. Phys.
                 {\bf B365}, 138 (1991); \cr

\+     &  See also  references listed in these articles.\cr

\medskip

\+ [6] & R.~Griganis, P.~J.~O'Donnell, M.~Sutherland and H.~Navalet,
                  Phys. Lett. {\bf 213 B}, \cr
\+     & 355 (1988); R.~Griganis, P.~J.~O'Donnell and M.~Sutherland,
          Phys. Lett. {\bf 237 B},  \cr
\+     & 252 (1990); P.~J.~O'Donnell and H.~K.~K.~Tung,  Phys. Rev.
                 {\bf D 45}, 252 (1990); \cr
\+     & G.~Cella, G.~Curci, G.~Ricciardi and A. Vicer\'e,  Phys. Lett.
            {\bf 248 B}, 181 (1990). \cr

\medskip

\+ [7] & E. Thorndike, CLEO Collaboration, talk given at the {\it 1993
        Meeting of the American }\cr
\+     & {\it  Physical Society, Washington, D. C. April,
        1993} . \cr

\medskip

\+[8] & Schoonschip program by M.~Veltman, unpublished. \cr

\medskip

\+ [9] & W.~Zimmermann, in {\sl Lectures in Elementary Particles and Quantum
        Field} \cr
\+ &  {\sl Theory}, edited by S.~Deser {\sl et al.}
    (MIT Press, Cambridge, Mass.,
      1971),\cr
\+ & Vol. I, p.397;
Y.~Kazama and  Y.-P.~Yao,  Phys. Rev. {\bf D 25}, 1605 (1982).\cr

\medskip

\+ [10] & T.~Inami and C.~S.~Lim,  Progr. Theor. Phys.  {\bf 65}
         297 (1981). \cr

\medskip

\+ [11] & N.~Cabibbo and L.~Maiani,  Phys. Lett.
                 {\bf 79 B}, 109 (1978); \cr
\+     & B.~A.~Campbell and P.~J.~O'Donnell,  Phys. Rev.
               {\bf D 25}, 1989 (1982). \cr
\medskip

\+  [12] & J. van der Bij and M. Veltman, Nucl. Phys. {\bf B231},
           205 (1984). \cr
\vfill \eject

\noindent
{\it FIGURE CAPTION:}
\bigskip
\noindent
Figure 1: $\bar{C}_{O_{52}}(m_b)$ dependence on $m_t$ to order $\alpha_s$
with $m_b = 4.8\;GeV$, $m_w=81\;GeV$, $\alpha_s(m_b)=0.19$.
See text for explanation of various curves.

\smallskip \noindent
Figure 2: Branching ratio for $b\rightarrow s \gamma$
as a function of $m_t$ with $QCD$ corrections. The solid line represents
the interpolation given by Eq. $(8.a)$; the dashed line represents
the interpolation given by Eq. $(8.b)$; the dotted line represents
the values of the branching ratio for $m_t = 140 \ GeV$.
We used $m_b = 4.8\;GeV$, $m_w=81\;GeV$, $\alpha_s(m_b)=0.19$.

\end

\vfill \eject

\input epsf
\epsfxsize=\hsize

\vfill
\epsffile{/Net/christie/usr/kassa/ci_2loop/BSG/Mt_approx_Mw/BSA/C52.eps}
\vfill

\centerline{ Figure 1.}
\vfill \eject

\epsfxsize=\hsize
\vfill
\epsffile{/Net/christie/usr/kassa/ci_2loop/BSG/Mt_approx_Mw/BSA/Branching_bsA.eps}
\vfill

\centerline{ Figure 2.}
\vfill \eject
\end